\documentclass[aps, pre, floatfix,
reprint,groupedaddress,
showpacs
]{revtex4-1}

\usepackage{graphicx}
\usepackage{bm}

\usepackage{amsmath}
\usepackage{amssymb}
\usepackage{color}
\usepackage{subfigure}
\usepackage{times}
\usepackage{float}
\usepackage[sort&compress]{natbib}

\usepackage[colorinlistoftodos,prependcaption,textsize=small]{todonotes}

\usepackage{hyperref}

\hypersetup{
  colorlinks=true,
citecolor=black,
linkcolor=black,
urlcolor=black
  }

\usepackage[T1]{fontenc}
\usepackage[utf8]{inputenc}

\newcommand{\bdt}[1]{{\color{orange} #1}}
\renewcommand{\bdt}[1]{{\color{blue} #1}}

\newcommand{\lopt}{l_{\mathrm{opt}}}

\renewcommand{\bdt}[1]{{\color{black} #1}}

\begin{document}

\title{Physics of free climbing}

\author{Bart{\l}omiej Dybiec}
\email{bartek@th.if.uj.edu.pl}
\affiliation{Institute of Theoretical Physics, Department of Statistical Physics, Jagiellonian University, \L{}ojasiewicza 11, 30-348 Krak\'ow, Poland}

\author{Karol Capa{\l}a}
\email{karol@th.if.uj.edu.pl}
\affiliation{Institute of Theoretical Physics, Department of Statistical Physics, Jagiellonian University, \L{}ojasiewicza 11, 30-348 Krak\'ow, Poland}

\author{Jakub Barbasz}
\email{ncbarbas@cyf-kr.edu.pl}
\affiliation{J. Haber Institute of Catalysis and Surface Chemistry, Polish Academy of Sciences, Niezapominajek 8, 30-239 Krak\'ow, Poland}


\date{\today}

\begin{abstract}
Theory of stochastic processes provides theoretical tools which can be efficiently used to explore properties of noise induced escape kinetics.
Since noise facilitated escape over the potential barrier resembles free climbing, one can use the first passage time theory in analysis of rock climbing.
We perform the analysis of the mean first passage time in order to answer the question regarding the optimal, i.e., resulting in the fastest climbing, rope length.
It is demonstrated that there is a discrete set of favorable rope lengths assuring shortest climbing times, as they correspond to local minima of mean first passage time.
Within the set of favorable rope lengths there is the optimal rope giving rise to the shortest climbing time.
In particular, more experienced climbers can decrease their climbing time by using longer ropes.
\end{abstract}


%
\maketitle

\setlength{\tabcolsep}{0pt}

%
%
\section{Introduction}

Noise induced escapes from potential wells as well as noise driven transitions over potential barriers have been studied for a long time.
Pioneering works of Farkas \cite{farkas1927} and Kramers \cite{kramers1940} related the escape rate to the height of the potential barrier and the system temperature.
Further development in the theory of stochastic process and Brownian motion \cite{borodin2002} allow for detailed studies of noise driven systems on theoretical \cite{mcnamara1989,hanggi1990,gammaitoni1998,doering1992}  and experimental \cite{russell1999use,evstigneev2005,schmitt2006} levels.
Presence of noise can facilitate escape kinetics resulting in optimal input/output synchronization (stochastic resonance \cite{mcnamara1989}), fastest escape kinetics (stochastic resonant activation \cite{doering1992}) and directed transport (Brownian motors \cite{magnasco1993,reimann2002}) to name a few.
These effects turned out to be relevant in practical applications \cite{simonotto1997visual,priplata2002noise} and biological realms \cite{goelrichter1974,hanggi2002}.
Properties of stochastic systems can be significantly affected by stochastic resetting \cite{evans2011diffusion,pal2019first,evans2020stochastic}, which at selected  time instants brings a particle back to a given point.
The restarting captures real life phenomena like returning to the home range, starting from the scratch or starting a new attempt.

\bdt{
The noise induced barrier crossing is a process underlying many physical phenomena.
The efficiency of the barrier crossing can be characterized by the mean first passage time (MFPT), which measures the average time needed to cross the barrier for the first time.
The MFPT  can be optimized in various manners but only some of them are applicable in the context of rock climbing, which we intend to model here.
For example in the stochastic resonant activation simultaneous action of noise and modulation of the barrier can expedite the escape kinetics \cite{doering1992}.
Analogous function plays stochastic resetting which eliminates subotimal trajectories \cite{evans2020stochastic}.
The MFPT can also be optimized by shaping \cite{li2017transports,li2016levy,li2020transition,innerbichler2020enhancing,chupeau2020optimizing}, e.g., corrugating the potential barrier, under action of Gaussian \cite{innerbichler2020enhancing,chupeau2020optimizing} and non-Gaussian noises \cite{li2017transports,li2016levy,li2020transition}.
In these setups the system is typically one dimensional and the motion is defined by setting the initial and final points.
The optimization of the MFPT in 1D does not affect the route, in the sense that all intermediate points belong to the interval $(x_i,x_f)$.
On the one hand, rock climbing and hiking can be performed along existing routes, which are almost 1D.
On the other hand, in order to expedite the time needed to reach the final point, one can relax the assumption that the trajectory is fixed by allowing the hiker/climber to select not only grips but also intermediate points within a large range.
Such a scenario moves the optimization problem to the graph-theory \cite{west2001introduction,bollobas2013modern} dealing, among others, with problems of finding shortest paths \cite{beardwood1959shortest,frank1969shortest,fu1998expected}, average path length \cite{fronczak2004average} and the traveling salesman problem \cite{applegate2011traveling,laporte1990selective}.
Obviously, a more general framework gives more space for the optimization as it provides the possibility to avoid more difficult parts of the route.
Nevertheless, it can violate the  rules of the ``climbing game'', because it could excessively reduce climbing difficulties.
For the sake of the climber's ethics, within the studied model, we assume that the climbing route is marked out with such precision that it allows only minimal freedom to choose climbing grips only.
}


Rock climbing is one of the processes which bears a lot of similarities with surmounting of the potential barrier in the noise-driven systems and stochastic resetting.
Following these links, we reinterpret the problem of optimal rope length \bdt{for motions along fixed paths}.
We assume that the climber acts as a noise driven particle while the rope assures that during the climbing a climber cannot come down below the beginning of a pitch.
Therefore, a beginning of each pitch serves as the reflecting boundary or a point to which a climber can be ``reset'' if he makes a mistake.
\bdt{This is consistent with the free climbing ethics, which requires a return to the starting point of a pitch after any strain on the belay system (i.e., falling off, hanging on the rope, catching a hook or a loop).}
The whole climbing route is divided into multiple pitches and each of them need to be surmounted.
In such a scenario one can ask several questions: Is there the optimal rope length which assures minimal climbing time? Does the optimal rope length depend on the climber's skills?
Using the concept of the mean first passage time, we optimize total climbing time with respect to the rope length.
We demonstrate that typically there exists the optimal rope length, which is the increasing function of the climber skills.
Consequently, more experienced climbers can use longer ropes and divide the whole route into a smaller number of longer pitches.

\bdt{
Hiking and rock climbing are inevitably connected with estimation of the time needed to complete the route.
For hiking this time can be predicted initially using, for example, the Naismith's rule \cite{naismith1892excursions}, which assumes 12 min per 1 km (on the map) and additional 10 minutes for every  100 m in ascent.
It is used, along with its extensions, in mountain guidebooks.
The Naismith's rule not only gives a reasonable estimate for the hiking time but it shows that not only the distance but also the slope matters.
The Naismith's rule can be further modified \cite{langmuir1969mountain,irmischer2018measuring}.
For longer routes one needs to incorporate the fatigue factor, which is introduced by the Tranter's correction.
Therefore, for sufficiently long routes, especially on inclined surfaces, the time is no longer proportional to the distance.
The fatigue factor is also observed in the other activities, e.g., running, as the time to finish a marathon is not equal to twice the time to complete a half-marathon.
Typically this time is a couple of percent larger than two.
The estimate can be provided by the Riegel's formula \cite{riegel1997time,blythe2016prediction} which states that the time to complete a marathon is $2^{1.06}\approx 2.085$ times longer that the time to finish half-marathon \footnote{Using data for the 18. PZU Cracovia Marathon (https://wyniki.datasport.pl/results2888/), on average, we get 2.09.}.
Consequently, already for movement along flat comfortable surfaces the time to complete the route does not need to be a linear function of the distance.
The situation gets more difficult on horizontal surfaces, which are not fully flat.
In that context, we can think for example about walking through an ice field that may be crisscrossed by crevasses.
The climber does not enter the crevasses (thus potential energy is constant), but must jump over them, overcome them with the help of a ladder or even bypass them.
Consequently, the motion is significantly disturbed and the mountaineer does not have to move at a constant speed.
The fatigue factor can be quite large and increases in time.  
Thus, the time of motion does not need to scale linearly with the distance.
}

\bdt{
In contrast to hiking, which actually is a long, brisk walk, climbing is the activity in which hands, feet and other parts of the body are used to climb steep objects.
During hiking the difficulty comes from: distance and ascent. During climbing, in addition to length and ascent, the difficulty (grade) of the climbing route is the  important factor.
The difficulty is further amplified by the tiredness, difficulties with proper regeneration \cite{heyman2009effects} and energy needs \cite{norman2004running}.
Consequently, the relation  between time and distance gets more complex.
Moreover, in hiking, chances of failure are smaller and after failure hikers typically do not take the second attempt right after the failure, while re-attempting is common for climbers.
Therefore, in overall, we consider the Brownian motion with restarting as a proxy to describe climbing, especially in multi pitch routes.
}






Similar considerations as described in the context of rock climbing can be carried out for Himalaya mountaineering, especially done in winter.
The risk of having to turn back between camps due to random factors or components independent of the climber (such as the weather) increases with the distance between camps and decreases with the climber's experience.
In the event of a withdrawal between bivouacs, the climber returns to the previous camp. Having more experience allows climbers to reduce the number of bivouacs by increasing the distance between them.
There is an optimal distance between the camps that allows climbers to overcome the wall in the shortest time.
Therefore, the described model should be treated as a universal model of climbing.
Finally, in the context of rock climbing and mountaineering one should mention Marian Smoluchowski for his contribution not only to the theory of Brownian motion \cite{smoluchowski1906b2,smoluchowski1916} and stochastic processes \cite{ulam1957marian} but also for  his achievements in mountain climbing \cite{fulinski1998apb}.


\section{Model}

The free climbing model is based on the overdamped Langevin equation \cite{risken1996fokker,gardiner2009}
\begin{equation}
    \frac{dx}{dt} = -V'(x) + \sigma \xi(t),
    \label{eq:langevin}
\end{equation}
which describes the 1D noise driven motion in the external potential $V(x)$. In Eq.~(\ref{eq:langevin}), $\xi(t)$ stands for the Gaussian white noise satisfying  $\langle \xi(t) \rangle=0$ and $\langle \xi(t) \xi(s) \rangle=\delta(t-s)$.
For a stochastic process one can study the first passage time problems \cite{gardiner2009}.
For a particle starting at $x=x_0$, the mean exit time (mean first passage time) $\mathcal{T}(x_0)$ from the bounded domain restricted by the reflecting boundary at $x=A$ and absorbing boundary at $x=B$ ($A<B$) \cite{gardiner2009} reads
\begin{equation}
    \mathcal{T}(x_0 \to B) =  \frac{2}{\sigma^2}\int_{x_0}^B dy \exp\left[ \frac{2V(y)}{\sigma^2} \right]
    \int_{A}^{y} dz \exp\left[ -\frac{2V(z)}{\sigma^2} \right].
    \label{eq:mfpt}
\end{equation}
The problem of optimization of climbing time will be explored by analyzing the mean first passage time in systems with \bdt{moving} boundaries.
In other words, if the climber finishes a pitch, reflecting and absorbing boundaries are moved to new positions.
\bdt{The reflecting boundary replaces the end of the completed pitch, while the absorbing boundary is moved to the most distant (up the hill) point which can be reached with the used rope.}
The heart of the theoretical framework is provided by Eq.~(\ref{eq:mfpt}) as it allows us to calculate the average time to make a pitch.

\begin{figure}[!h]%
\centering
\begin{tabular}{c}
\includegraphics[angle=0,width=0.7\columnwidth]{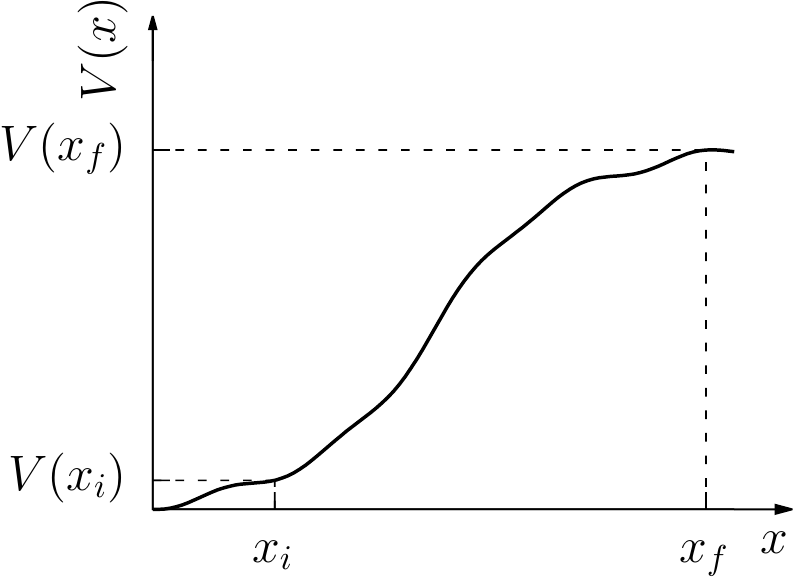} \\
\end{tabular}
\caption{A sample potential profile $V(x)$ representing the climbing route. $x_i$ and $x_f$ stand for starting (reflecting) and final (absorbing) positions.
The path is characterized by the width (distance between boundaries) $\Delta=x_f-x_i$, height difference $H=V(x_f)-V(x_i)$ and the total length $L=\int_{x_i}^{x_f}\sqrt{1+[V'(x)]^2}dx$.
During climbing the whole path is divided into pitches of a fixed length (\bdt{measured along the slope})  determined by the rope length $l$.
}
\label{fig:potential}
\end{figure}

For the fixed potential $V(x)$, using initial $x_i$ and final $x_f$ positions, see Fig.~\ref{fig:potential}, from Eq.~(\ref{eq:mfpt}) one can  calculate the mean time which is required to climb any point, e.g., the top (maximum of the potential).
The path is characterized by its total length $L=\int_{x_i}^{x_f} \sqrt{1+[V'(x)]^2}dx$, width $\Delta = x_f-x_i$ and height difference $H=V(x_f)-V(x_i)$.
In real situations, $V(x)$ is typically non-decaying function on $[x_i,x_f]$, i.e., $V'(x) \geqslant 0$ for $x\in [x_i,x_f]$.
\bdt{Nevertheless, it might also happen that $V(x)$ decays.
In such a case the climber can perform  abseiling, free solo descending or free descending.
Among these option abseiling is the fast one while two others scenarios are disproportionately long relative to the MFPT predictions.
Consequently, the locally decaying $V(x)$ cannot be easily incorporated into the studied model and we leave this issue for further studies.
}
During climbing the whole path is divided into multiple pitches, which are assumed to be of a fixed (\bdt{measured along the slope})  length determined by the rope length $l$.
Only the last pitch can be shorter than the remaining ones.

The process of surmounting the potential in the presence of reflecting and absorbing boundaries resembles climbing.
The first reflecting boundary ($x_i$) is the base for attacking the wall, while the ultimate absorbing boundary is the point to be reached ($x_f$).
The MFPT calculated directly from Eq.~(\ref{eq:mfpt}) assumes that the whole potential barrier is traversed at once.
Nevertheless, for each trajectory, (multiple) returns to the base point $x_i$ and \bdt{revisits to other points} are allowed \bdt{as stochastic trajectories meander between absorbing and reflecting boundaries}.
In free climbing, climbers use climbing equipment, e.g., ropes, to protect against injury during falls and to limit the length of segments (pitches).
Ends of successfully passed segments (determined by the full rope length $l$) are used as new starting points, which can be considered as further reflecting boundaries.
In course of time, the reflecting boundary shifts upwards by discrete steps, consequently after the unsuccessful attempt the climber starts not from the base point $x_i$ but from the final point of the previous segment.
This effect is the analog of stochastic resetting \cite{evans2011diffusion,evans2011diffusion-jpa,evans2020stochastic}, but the restart (\bdt{to the end of the last competed pitch}) takes place after each mistake.
The mechanism of \bdt{moving} reflecting and absorbing boundaries significantly changes the barrier surmounting process.
In this context, one can ask the question what is the optimal rope length $l$ which gives the smallest time to reach the final point, e.g., the top of the mountain $x_f$, see Fig.~\ref{fig:potential}.
The optimal length $l$ needs to optimize the full time needed to reach the top of the wall, which consists of climbing time (including the time needed to secure the rope) and constant time needed to prepare the starting point for every pitch.
The whole route is divided into many segments (pitches).
Each segment starts with a reflecting boundary ($A$) and ends with the absorbing one ($B$).
We assume that time needed to pass each segment is the sum of the climbing time given by Eq.~(\ref{eq:mfpt}) and time to prepare every pitch $\kappa$.
For simplicity, it is assumed that $\kappa$ is a constant and independent of the rope length $l$.
Nevertheless, we have also checked other options, which have not changed results qualitatively, proving generality of the drawn conclusions.
Finally, the MFPT depends on the system temperature, i.e., $\sigma=\sqrt{2T}$.
The $\sigma$ parameter in Eq.~(\ref{eq:langevin}) measures the climber skills.
The larger $\sigma$ the better climber, i.e., the MFPT decays with the increase in $\sigma$, see Eq.~(\ref{eq:mfpt}).

%
%
\section{Results \& Discussion}

We use fully solvable cases of free motion ($V(x)=0$) and constant slope ($V(x) \propto x$) which are already capable of revealing general properties of the free climbing model.
Afterwards, we switch to a parabolic wall ($V(x) \propto x^2$) and discuss the general case.

%
\subsection{Flat horizontal wall (free motion)}

\begin{figure}[!h]%
\centering
\begin{tabular}{c}
\includegraphics[angle=0,width=0.7\columnwidth]{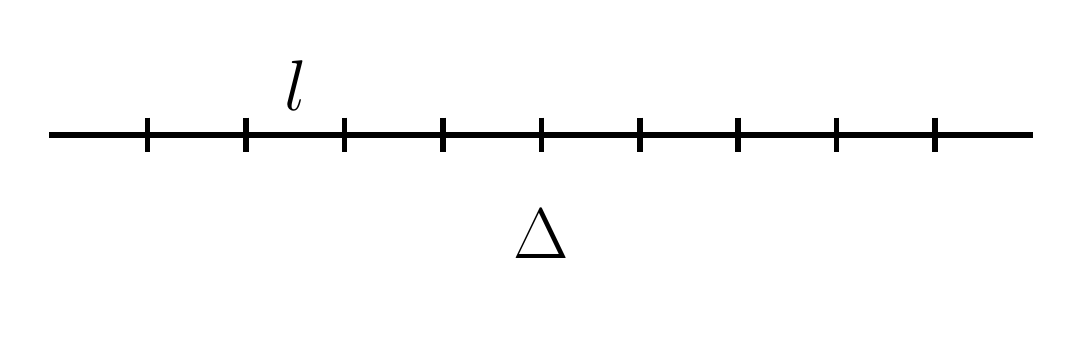} \\
\end{tabular}
\caption{A finite interval of length $L$ and width $\Delta$ ($\Delta = L$) is divided into $\Delta/l$ segments of length $l$.
}
\label{fig:interval}
\end{figure}

We start with the simplest case of $V(x)=0$, see Fig.~\ref{fig:interval}.
In such a case the distance between boundaries ($\Delta$) and route length ($L$) are the same.
For $A=0$   Eq.~(\ref{eq:mfpt}) gives
\begin{equation}
    \mathcal{T}(x_0 \to B) = \frac{(B -x_0)(B +x_0)}{\sigma^2}.
    \label{eq:mfpt-flat}
\end{equation}
\bdt{As follows from Eq.~(\ref{eq:mfpt-flat}) the time needed to pass the given distance does not scale linearly with the distance.
One can think about this lack of linearity as the fatigue factor which can be further amplified by the various types of obstacles on the flat surface, e.g., on an ice field.
}

The climber starts the climbing at the reflecting boundary, i.e., $x_i=0$, and the whole interval $\Delta$ is divided into segments given by the rope length $l$.
Each segment can be passed in the same, fixed time $\mathcal{T}(0 \to l)+\kappa$, i.e.,
$\frac{l^2}{\sigma^2} + \kappa$, because in the absence of the deterministic force all pitches are the same.
In the model (Model~A), the time needed to pass the whole route $\Delta$ is equal to
\begin{equation}
\mathcal{T} =   \frac{\Delta}{l}  \left[  \frac{l^2}{\sigma^2} + \kappa \right].
\label{eq:free-cont}
\end{equation}
\bdt{The Model~A, defined by Eq.~(\ref{eq:free-cont}), can be also called the fractional/partial model, because it incorporates only a fraction of time $\kappa$ to prepare the last pitch.
}
The time given by Eq.~(\ref{eq:free-cont}) is minimal for
\begin{equation}
    \lopt=\min\{\sigma \sqrt{\kappa} , \Delta\}.
\end{equation}
If the rope length is smaller than the interval width, i.e., $\lopt= \sigma \sqrt{\kappa} < \Delta$, the mean exit time is equal to
\begin{equation}
    \mathcal{T}_{\mathrm{min}}= \frac{2 \sqrt{\kappa } \Delta }{\sigma }.
\end{equation}
Otherwise, optimal rope length is $\lopt=\Delta$ and $ \mathcal{T}_{\mathrm{min}}=\frac{\Delta^2}{\sigma^2}+\kappa$.
For $\lopt=\Delta$, the further increase in the rope length does not change the climbing time.
The mean exit time grows with increasing $L$ ($\Delta$) (longer overall distance), increasing securing time $\kappa$ (longer breaks) and decreasing $\sigma$ (decreasing climbers’ skills).
In the limit of $\kappa\to 0$, the climbing time vanishes because time to pass each segment tends to zero as $l^2$, while the number of segments grows like $1/l$ making the product behave as $l$.
\bdt{In this very special, artificial limit, both the climbing time and the optimal rope length tend to 0.
Consequently, the model reveals unphysical behavior as velocity growths to infinity, i.e., it is greater than the speed of light ($v \gg c$).
On the one hand, this controversy seems to be apparent, because of the unrealistic assumption of $\kappa\to0$, meaning that a climber can start a new pitch in zero time.
On the other hand, for finite $\kappa$ and rope length $l\to 0$ the climbing time diverges as the climber needs to start an infinite number of pitches.
Nevertheless, already the $l\to 0$ limit is problematic, because the climber moves forward due to shifting of the reflecting boundary.
}

\begin{figure}[!h]%
\centering
\begin{tabular}{c}
\includegraphics[angle=0,width=0.98\columnwidth]{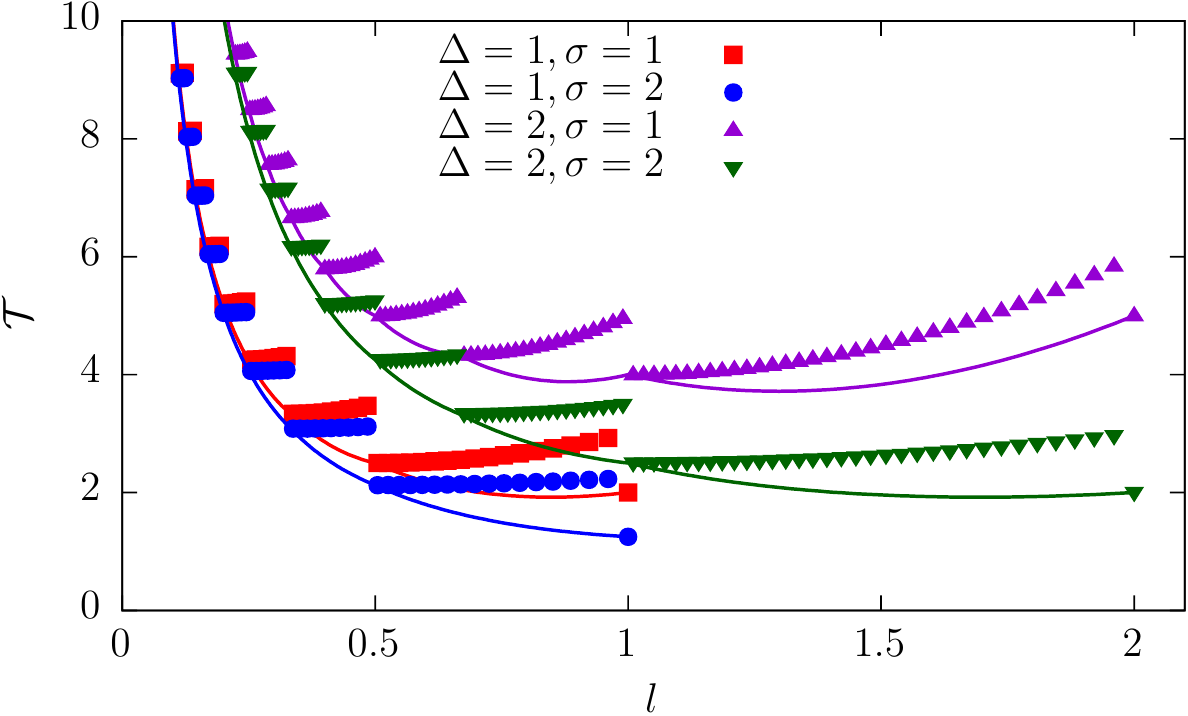} \\
\end{tabular}
\caption{(Color online): The dependence of the MFPT on the rope length $l$ for a free particle with $\Delta=1$ and $\Delta = 2$ for $\kappa=1$.
Solid lines show results for the \bdt{Model~A (fractional/partial)}, see Eq.~(\ref{eq:free-cont}), while points to the \bdt{Model~B (discrete)}, see Eq.~(\ref{eq:free-disc}).
}
\label{fig:free}
\end{figure}

Formally one should consider the discrete (Model~B) version of the \bdt{fractional/partial} model, see Eq.~(\ref{eq:free-cont}), for which
\begin{equation}
\mathcal{T} =  \left\lfloor \frac{\Delta}{l} \right\rfloor    \left[  \frac{l^2}{\sigma^2} + \kappa \right]  + \frac{\Lambda^2}{\sigma^2} + \kappa,
\label{eq:free-disc}
\end{equation}
where $\Lambda=\Delta-\lfloor  \Delta/l \rfloor l$ is the length of the last pitch for which the whole time $\kappa$ has to be added to the overall climbing time.
In Eq.~(\ref{eq:free-disc}), the $\lfloor \Delta/l \rfloor$ stands for the floor function which returns the greatest integer less than or equal to $\Delta/l$, which is the number of pitches of length equal to the full rope length $l$.

In Fig.~\ref{fig:free} MFPTs as a function of rope length $l$ for various $\Delta$ and $\sigma$ are presented.
Solid lines represent the \bdt{fractional/partial} model, see Eq.~(\ref{eq:free-cont}) while points correspond to Eq.~(\ref{eq:free-disc}).
The main difference between \bdt{fractional/partial} and discrete models comes from the last pitch.
In both cases the last pitch is climbed at the same time but in the \bdt{fractional/partial} model only a fraction of $\kappa$ proportional to the length of the last pitch is added to the overall climbing time.

From the discrete model it is clearly visible that there is no benefit in increasing the rope length if it does not result in the reduction of the number of pitches. Therefore, considered rope length should satisfy $l=\Delta/n$, where $n \in \mathbb{N}$.
For intermediate rope lengths the number of pitches is unaffected, but the total passing time is increased, because pitches are longer.
For very experienced climbers it can turn out that the optimal rope length $l$ can be equal to the route length $L$.
Finally, the climbing time can be decreased by giving up the rope, i.e., in the free solo climbing.

Interesting situation is observed for the \bdt{fractional/partial} model.
For such a model the time needed to pass the whole route can be not only a non-monotonous function of the rope length, but it can be piecewise smooth. This is especially well visible for $\Delta=2$ with $\sigma=1$, where there exists such MFPT which can be recorded for three distinct values of the rope length.
Moreover, at favorable rope lengths, i.e., for $l=L/n$, the \bdt{fractional/partial} and the discrete models are equivalent, see Eqs.~(\ref{eq:free-cont}) and (\ref{eq:free-disc}).

%
\subsection{Fixed (linear) slope (linear potential)}

\begin{figure}[!h]%
\centering
\begin{tabular}{c}
\includegraphics[angle=0,width=0.7\columnwidth]{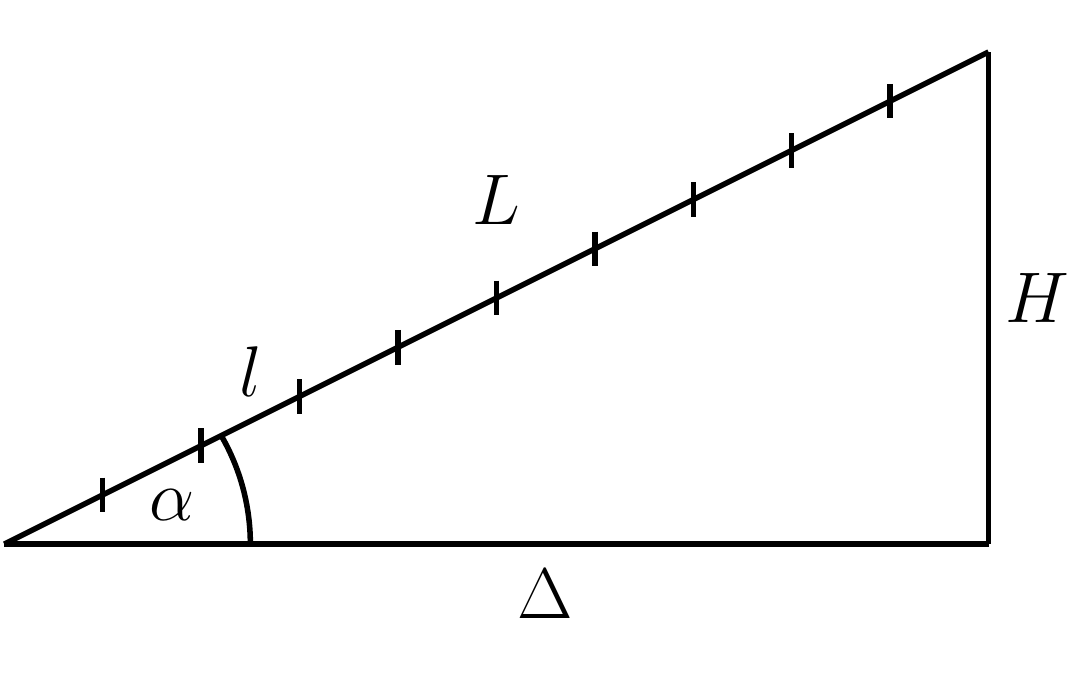} \\
\end{tabular}
\caption{The fixed linear slope $V(x)=\tan(\alpha)  x$ of total length $L=\Delta/\cos\alpha=H/\sin\alpha$ is divided into $(\Delta/\cos\alpha) / l$ segments of length $l$.
The slope of the potential is determined by the route parameters $\Delta$ and $H$, i.e., $\tan\alpha={H}/{\Delta}$.
}
\label{fig:slope-setup}
\end{figure}

The next fully tractable case is $V(x)=\tan(\alpha)  x$, which corresponds to the fixed linear slope with $\tan{\alpha}=H/\Delta$, see Fig.~\ref{fig:slope-setup}.
For $A=0$,  Eq.~(\ref{eq:mfpt}) gives the climbing time
\begin{equation}
    \mathcal{T}(x_0 \to B) = \frac{\sigma ^2 e^{\frac{2 B \tan\alpha}{\sigma ^2}}}{2 \tan^2\alpha}-\frac{B}{\tan\alpha}-\frac{\sigma ^2 e^{\frac{2 x_0 \tan\alpha}{\sigma ^2}}}{2 \tan^2\alpha}+\frac{x_0}{\tan\alpha}
\end{equation}
which for $x_0=0$ reduces to
\begin{equation}
    \mathcal{T}(0 \to B) = \frac{\sigma ^2 e^{\frac{2 B \tan\alpha}{\sigma ^2}}}{2 \tan^2\alpha}-\frac{B}{\tan\alpha}-\frac{\sigma ^2}{2 \tan^2\alpha}.
    \label{eq:t0slope}
\end{equation}
For each segment $B$ is determined by the rope length~$l$, i.e.,
$
    B={l}{\cos\alpha},
$
where $\alpha$ is the angle characterizing the slope ($\tan \alpha=H/\Delta$).
Analogously like for $V(x)=0$ the time to climb each segment is constant because of the constant slope (constant force).
In the \bdt{fractional/partial} model, the time needed to climb to the top is equal to
\begin{equation}
    \mathcal{T}=\frac{\frac{\Delta}{\cos\alpha }}{l} \left[  \mathcal{T}(0 \to l\cos\alpha) + \kappa  \right],
    \label{eq:slope-cont}
\end{equation}
where $\mathcal{T}(0 \to l\cos\alpha)$ is given by Eq.~(\ref{eq:t0slope}) with $B=l\cos\alpha$.
For ropes longer than the slope length, $l>L=\Delta/\cos\alpha = H/\sin\alpha$, the climbing time is equal to $\mathcal{T}(0\to\Delta)+\kappa$, where $\mathcal{T}(0\to\Delta)$ is given by Eq.~(\ref{eq:t0slope}) with $B=\Delta$.
In the limit of $\alpha\to 0$ the free motion, already considered in Fig.~\ref{fig:free}, is recovered.

\begin{figure}[!h]%
\centering
\begin{tabular}{c}
\includegraphics[angle=0,width=0.98\columnwidth]{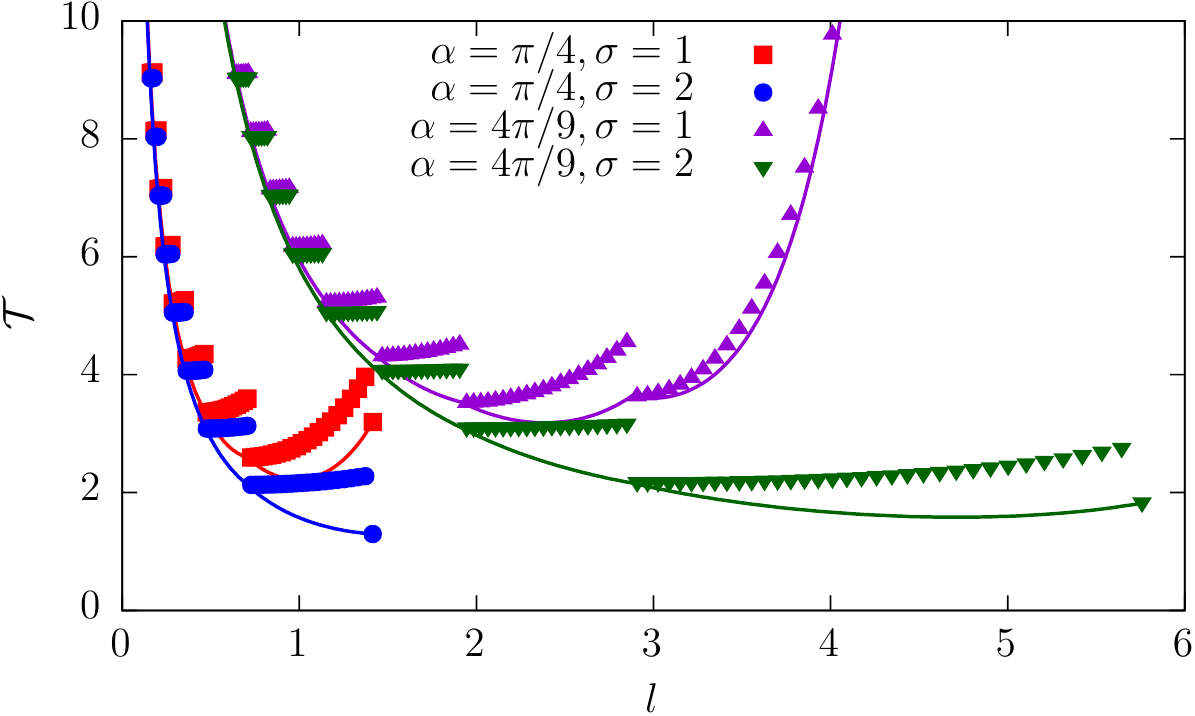} \\
\end{tabular}
\caption{(Color online): The dependence of the MFPT on the rope length $l$ for fixed slope with $\Delta=1$ and $\kappa=1$.
Various groups of curves correspond to various slopes of $\pi/4$ and $4\pi/9$.
Solid lines show results for the \bdt{Model~A (fractional/partial)}, see Eq.~(\ref{eq:slope-cont}), while points for the \bdt{Model~B (discrete)}, see Eq.~(\ref{eq:slope-disc}).
}
\label{fig:slope}
\end{figure}

Similarly like for a free walker, also for the fixed slope one should consider the discretized version
\begin{equation}
\mathcal{T}=\left\lfloor\frac{\frac{\Delta}{\cos\alpha }}{l}\right\rfloor \left[  \mathcal{T}(0 \to l\cos\alpha) + \kappa  \right] + \mathcal{T}(0\to\Lambda\cos\alpha)+\kappa,
\label{eq:slope-disc}
\end{equation}
where $\Lambda=\Delta/\cos\alpha-\lfloor  \Delta/l\cos\alpha \rfloor l$ the length of the last pitch.
For the constant slope, as for free motion, there is a discrete spectrum of favorable rope lengths $l=L/n$ ($n\in\mathbb{N}$).
Nevertheless, this time local minima of MFPT are better pronounced than for the free motion.
Fig.~\ref{fig:slope} depicts results for $\Delta=1$ with the fixed slope of $\alpha=\pi/4$  ($L=\Delta/\cos\alpha=\sqrt{2}$) and $\alpha=4\pi/9$  ($L=\Delta/\cos\alpha \approx 5.75$).

With the increasing time $\kappa$ the optimal rope length $\lopt$ (as long as it is smaller than the route length) increases, because the $\kappa/l$ term moves optimum to the right.
For growing $\alpha$ (with fixed $\Delta$) the route length $L$ and, as well as optimal rope length $\lopt$ grow.

%
\subsection{General walls (parabolic and more general potentials)}

\begin{figure}[!h]%
\centering
\begin{tabular}{c}
\includegraphics[angle=0,width=0.98\columnwidth]{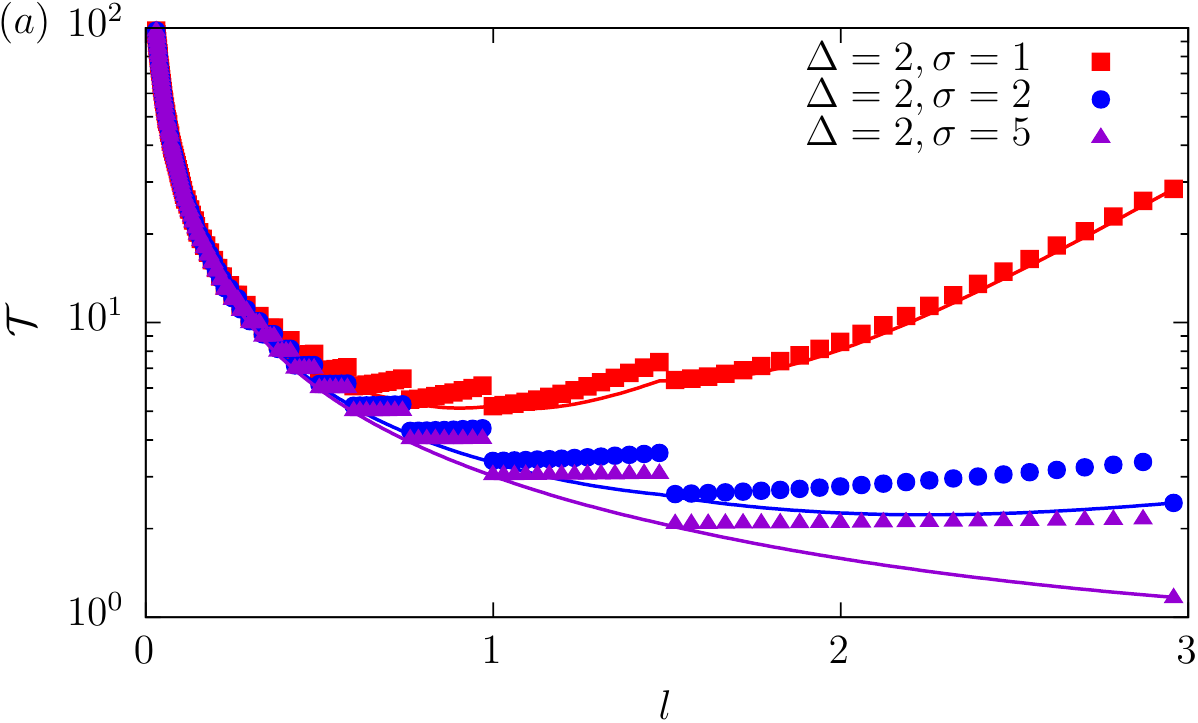} \\
\includegraphics[angle=0,width=0.98\columnwidth]{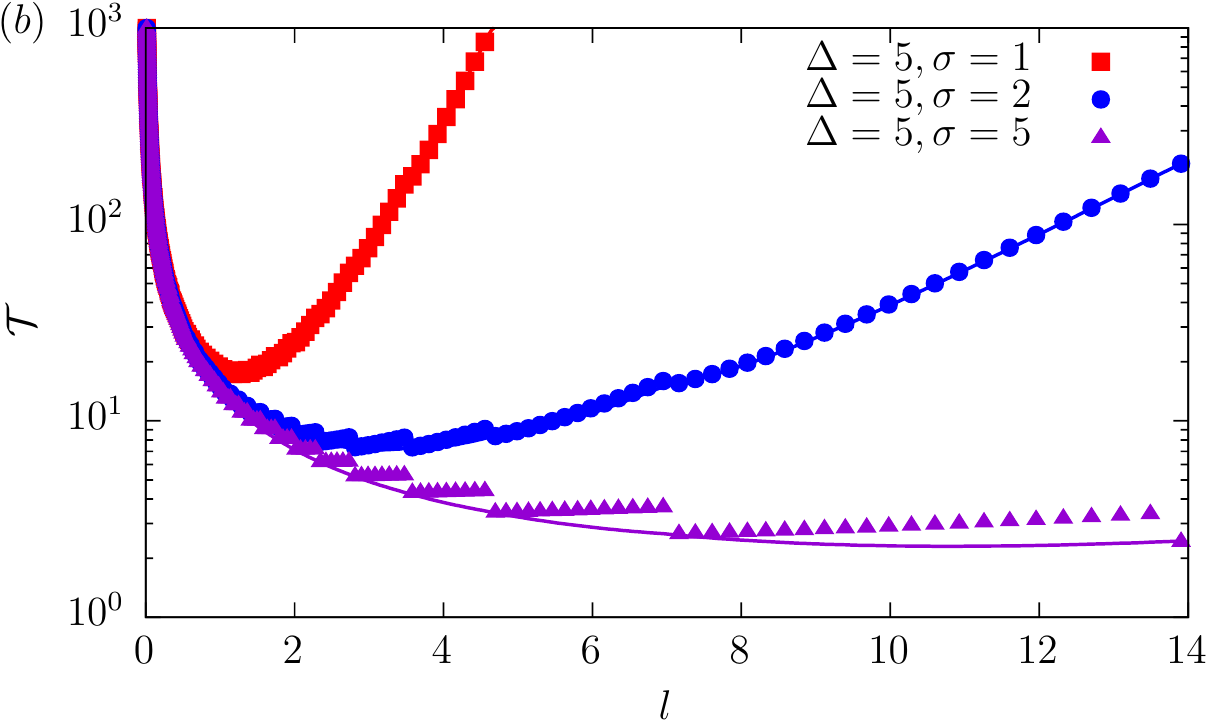} \\
\end{tabular}
\caption{(Color online): The dependence of the MFPT on the rope length $l$ for parabolic potential $V(x)=x^2/2$ with $\kappa=1$ and various widths $\Delta$.
Solid lines show results for the \bdt{Model~A (fractional/partial)}, while points for the \bdt{Model~B (discrete)}.
}
\label{fig:x2}
\end{figure}

For general potentials $V(x)$ the MFPT can be calculated by Eq.~(\ref{eq:mfpt}).
The calculation of MFPT corresponding to a given rope length $l$ is more complex, as widths and heights of the segment are determined by the rope length.
In particular, knowing the starting point $a$ and the rope length $l$ one can find the intermediate point $b$ to be reached from the following constraint
\begin{equation}
    l=\int_{a}^b \sqrt{1+[V'(x)]^2}dx.
\end{equation}
For the general potential $V(x)$, the very same, described above, resonant effect is also recorded.
Approximately the MFPT from $a \to b$ is proportional to $\exp[h/\sigma^2]$ with $h=V(b)-V(a)\approx (b-a) V'(a)$ \bdt{being the barrier height between $a$ and $b$}.
Therefore, the MFPT is the polynomial in $(b-a)$ and the total MFPT can be optimized with respect to $l$ in analogous ways like for a free particle and the fixed slope.
For instance, Fig.~\ref{fig:x2} presents results for $V(x)=x^2/2$ with various $\Delta$ and $\sigma$.
For $\Delta=2$ the route length $L\approx 2.96$ while for $\Delta=5$ $L\approx 13.9$.
Solid lines correspond to the \bdt{fractional/partial} model while points to the discrete model.
For $l=L/n$ ($n\in\mathbb{N}$) both models are equivalent.

%

Finally, in more general realms one can relax the assumption that the time $\tau$ needed to start each pitch is constant.
If $\tau = \kappa + f(l)$, where $f(l)$ is the increasing function of the rope length, combs-like parts of MFPT curves move up, making minima more pronounced.
The constant part of the securing time $\kappa$ assures that there exists the optimal rope length giving rise to the minimal total climbing time, while $f(l)$ makes minima corresponding to favorable rope lengths deeper.

%
%
\section{Final remarks}

The theory of stochastic processes is a widely accepted part of statistical physics which is typically used for examination of multiple physical models under action of noise.
Here, we have demonstrated that the very same tools can be effectively used to describe the process of rock climbing.

During rock climbing, long climbing routes are divided into smaller pitches (segments).
Every pitch works as the segment restricted by the reflecting boundary (beginning of the pitch) and absorbing boundary (end of the pitch).
The pitch length cannot be longer than the rope length, but here we have additionally assumed that all pitches (except the last one) are of the same fixed length given by the rope length.
The time needed to pass the whole route can be calculated using the mean first passage time formalism.
The minimal passing time is determined by the interplay between pitch climbing time and time needed to start each pitch.
\bdt{Under non-realistic assumption,} that segments could be prepared without any time cost (penalty) the optimal rope length tends to 0.
Otherwise, due to constant time penalty, there could exist an optimal rope length which is shorter than the route length.
The optimal rope length is selected among favorable rope lengths, i.e., $L/n$ ($n\in \mathbb{N}$).
Intermediate rope length does not decrease climbing time.
Finally, experienced climbers can use longer ropes as for longer ropes the total number of pitches is smaller.

%
%
\section*{Acknowledgements}

This research was supported in part by PLGrid Infrastructure and by the National Science Center (Poland) grant 2018/31/N/ST2/00598.
JB acknowledges financial support of the statutory research fund of ICSC PAS.  

%
%
\section*{Data availability}
The data that support the findings of this study are available from the corresponding author (BD) upon reasonable request.

%
%

\def\url#1{}

\end{document}